\documentclass[12pt]{article}%
\usepackage{amsmath,latexsym}
\usepackage{graphicx}
\usepackage{amsmath}
\usepackage{amsfonts}
\usepackage{amssymb}%
\begin{document}

\title{{Thin-shell wormholes admitting conformal motions 
   in spacetimes of embedding class one}}
   \author{
Peter K. F. Kuhfittig*\\
\footnote{E-mail: kuhfitti@msoe.edu}
 \small Department of Mathematics, Milwaukee School of
Engineering,\\
\small Milwaukee, Wisconsin 53202-3109, USA}

\date{}
 \maketitle

\begin{abstract}\noindent
This paper discusses the feasibility of 
thin-shell wormholes in spacetimes of 
embedding class one admitting a one-parameter 
group of conformal motions.  It is shown that 
the surface energy density $\sigma$ is positive, 
while the surface pressure $\mathcal{P}$ is 
negative, resulting in $\sigma + \mathcal{P} <0$,
thereby signaling a violation of the null energy 
condition, a necessary condition for holding a 
wormhole open.  For a Morris-Thorne wormhole, 
matter that violates the null energy condition 
is referred to as ``exotic."  For the thin-shell 
wormholes in this paper, however, the violation 
has a physical explanation since it is a direct 
consequence of the embedding theory in conjunction 
with the assumption of conformal symmetry.  These 
properties avoid the need to hypothesize the 
existence of the highly problematical exotic 
matter.   \\

\noindent
\textbf{Keywords} \\
Thin-Shell Wormholes, Conformal 
   Symmetry, Embedding Class One, Exotic Matter
  
\end{abstract}

\section{Introduction}\label{S:Introduction}
Wormholes are handles or tunnels in spacetime 
connecting different regions of our Universe
or entirely different universes.  While there
had been some forerunners, macroscopic traversable 
wormholes were first discussed in detail by 
Morris and Thorne in 1988 \cite{MT88}.  The 
wormhole geometry is described by the following 
static and spherically symmetric line element 
\begin{equation}\label{E:line1}
ds^{2}=-e^{\nu(r)}dt^{2}+\frac{dr^2}{1-\frac{b(r)}{r}}
+r^{2}(d\theta^{2}+\text{sin}^{2}\theta\,d\phi^{2}),
\end{equation}
using units in which $c=G=1$.  Here
$\nu=\nu(r)$ is usually referred to as the
\emph{redshift function}, which must be
everywhere finite to prevent the occurrence 
of an event horizon.  The function $b=b(r)$ 
is called the \emph{shape function} since it
determines the spatial shape of the
wormhole when viewed, for example, in an 
embedding diagram.  The spherical surface 
$r=r_0$ is called the \emph{throat} of the 
wormhole.  At the throat, $b=b(r)$ must 
satisfy the following conditions:
$b(r_0)=r_0$, $b(r)<r$ for $r>r_0$, and
$b'(r_0)\le 1$, usually called the
\emph{flare-out condition}.  This condition
can only be satisfied by violating the null
energy condition (NEC), which states that 
for the stress-energy tensor $T_{\alpha\beta}$,
we must have
\begin{equation}\label{E:nullvectors}
  T_{\alpha\beta}k^{\alpha}k^{\beta}
  \ge 0
\end{equation}
for all null vectors $k^{\alpha}$.  For 
the outgoing null vector $(1,1,0,0)$, the 
violation becomes 
\begin{equation}
   T_{\alpha\beta}k^{\alpha}k^{\beta}=
      \rho +p_r<0.
\end{equation}
Here $T^t_{\phantom{tt}t}=-\rho$ is the
energy density, $T^r_{\phantom{rr}r}= p_r$
is the radial pressure, and
$T^\theta_{\phantom{\theta\theta}\theta}=
T^\phi_{\phantom{\phi\phi}\phi}=p_t$ is
the lateral (transverse) pressure.  For a 
Morris-Thorne wormhole, matter that violates 
the NEC is called ``exotic," a term borrowed 
from quantum field theory.

The purpose of this paper is to account for 
the problematical nature of exotic matter by 
studying the effects of conformal symmetry 
in conjunction with some well-known classical
embedding theorems.  More precisely, by 
conformal symmetry we mean the existence 
of a conformal Killing vector $\xi$ defined 
by the action of $\mathcal{L_{\xi}}$ on the 
metric tensor:
\begin{equation}
  \mathcal{L_{\xi}}g_{\mu\nu}=\psi(r)\,g_{\mu\nu};
\end{equation}
here $\mathcal{L_{\xi}}$ is the Lie derivative
operator and $\psi(r)$ is the conformal factor.
Embedding theorems, which have their origin in 
classical geometry, depend on Campbell's theorem, 
which has been used to show that a Riemannian 
space can be embedded in a higher-dimensional 
flat space.

\section{Conformal Killing vectors}
    \label{S:Killing}
As indicated in the Introduction, we assume in this 
paper that our static spherically symmetric spacetime 
admits a one-parameter group of conformal motions, 
by which we mean motions along which
the metric tensor of a spacetime remains invariant
up to a scale factor.  In other words, there exist
conformal  Killing vectors such that
\begin{equation}\label{E:Lie}
   \mathcal{L_{\xi}}g_{\mu\nu}=g_{\eta\nu}\,\xi^{\eta}
   _{\phantom{A};\mu}+g_{\mu\eta}\,\xi^{\eta}_{\phantom{A};
   \nu}=\psi(r)\,g_{\mu\nu},
\end{equation}
where the left-hand side is the Lie derivative of the
metric tensor and $\psi(r)$ is the conformal factor
\cite{MM96, BHL07}.  In the usual terminology, the 
vector $\xi$ generates the conformal symmetry and 
the metric tensor $g_{\mu\nu}$ is conformally mapped 
into itself along $\xi$.  According to Refs. 
\cite{HP85a, HP85b}, this type of symmetry has 
proved to be effective in describing relativistic 
stellar-type objects.  Furthermore, conformal 
symmetry has led to new solutions, as well 
as to new geometric and kinematical insights 
\cite{MS93, sR08, fR10, fR12}.  Two earlier 
studies assumed \emph{non-static} conformal 
symmetry \cite{BHL07, BHL08}.

To study the effect of conformal symmetry, we 
wish to make use of Ref. \cite{WP92}, which 
uses the following form of the line element:
\begin{equation}\label{E:line2}
   ds^2=- e^{\nu(r)} dt^2+e^{\lambda(r)} dr^2
   +r^2( d\theta^2+\text{sin}^2\theta\, d\phi^2).
\end{equation}
The Einstein field equations then become

\begin{equation}\label{E:Einstein1}
e^{-\lambda}
\left(\frac{\lambda^\prime}{r} - \frac{1}{r^2}
\right)+\frac{1}{r^2}= 8\pi \rho,
\end{equation}

\begin{equation}\label{E:Einstein2}
e^{-\lambda}
\left(\frac{1}{r^2}+\frac{\nu^\prime}{r}\right)-\frac{1}{r^2}=
8\pi p_r,
\end{equation}

\noindent and

\begin{equation}\label{E:Einstein3}
\frac{1}{2} e^{-\lambda} \left[\frac{1}{2}(\nu^\prime)^2+
\nu^{\prime\prime} -\frac{1}{2}\lambda^\prime\nu^\prime +
\frac{1}{r}({\nu^\prime- \lambda^\prime})\right] =8\pi p_t.
\end{equation}
Here $\rho$ is the energy density, $p_r$ is the radial 
pressure, and $p_t$ is the transverse pressure.  It is 
well known that Eq. (\ref{E:Einstein3}) could be obtained
from the conservation of the stress-energy tensor, i.e.,
$T^{\mu\nu}_{\phantom{\mu\nu};\nu}=0$.  So we need to use
only Eqs. (\ref{E:Einstein1}) and (\ref{E:Einstein2}).

As pointed out by Herrera and Ponce de Le\'{o}n 
\cite{HP85a}, the subsequent analysis can be simplified 
somewhat by restricting the vector field in a certain
way: we require that $\xi^{\alpha}U_{\alpha}=0$, where 
$U_{\alpha}$ is the four-velocity of the perfect fluid
distribution, and that fluid flow lines are mapped 
conformally onto fluid flow lines.  According to Ref. 
\cite{HP85a}, the assumption of spherical symmetry 
then implies that $\xi^0=\xi^2=\xi^3=0$.  Eq. 
(\ref {E:Lie}) now yields the following results:
\begin{equation}\label{E:sol1}
    \xi^1 \nu^\prime =\psi,
\end{equation}
\begin{equation}\label{E:sol2}
   \xi^1  = \frac{1}{2}\psi r,
\end{equation}
and
\begin{equation}\label{E:sol3}
  \xi^1 \lambda ^\prime+2\,\xi^1 _{\phantom{1},1}=\psi.
\end{equation}
From Eqs. (\ref{E:sol1}) and (\ref{E:sol2}), we then
obtain
\begin{equation} \label{E:gtt}
   e^\nu  =C r^2,
\end{equation}
where $C$ is an integration constant.  Combined with 
Eq. (\ref{E:sol3}), this yields
\begin{equation}\label{E:grr}
   e^\lambda  = \left(\frac {1} {\psi}\right)^2.
\end{equation}

The arbitrary constant in Eq. (\ref{E:gtt}) 
can be obtained from the junction conditions 
in the usual way.  This is a necessary step 
since, according to Eq. (\ref{E:gtt}), our 
wormhole spacetime is not asymptotically flat 
and must therefore be cut off at some $r=a$ 
and joined to an exterior Schwarzschild 
spacetime,
\begin{equation}
ds^{2}=-\left(1-\frac{2M}{r}\right)dt^{2}
+\frac{dr^2}{1-2M/r}
+r^{2}(d\theta^{2}+\text{sin}^{2}\theta\,
d\phi^{2}).
\end{equation}
It follows that $e^{\nu(a)}=Ca^2=1-2M/a$, 
so that
\begin{equation}\label{E:cutoff}
   C=\frac{1-2M/a}{a^2},
\end{equation}
where $M$ is the mass of the wormhole as seen
by a distant observer.  We also have $b(a)=2M$.

For future reference, let us note that the field 
equations (\ref{E:Einstein1}) and (\ref{E:Einstein2}) 
can be rewritten as follows:
\begin{equation}\label{E:E1}
\frac{1}{r^2}(1 - \psi^2)-
\frac{(\psi^2)^\prime}{r}= 8\pi \rho
\end{equation}
and
\begin{equation}\label{E:E2}
\frac{1}{r^2}(3\psi^2-1)= 8\pi p_r.
\end{equation}

To see why, we get from Eq. (\ref{E:grr})
\begin{equation*}
   e^{-\lambda}=\psi^2\quad\text{and}\quad
   \lambda'=-\frac{2\psi'}{\psi}.
\end{equation*}
Substituting in Eq. (\ref{E:Einstein1}), 
we get
\begin{equation*}
   \psi^2\left(-\frac{2\psi'}{\psi r}
   -\frac{1}{r^2}\right)+\frac{1}{r^2}=
   -\frac{2\psi\psi'}{r}-\frac{1}{r^2}\psi^2
   +\frac{1}{r^2}=\frac{1}{r^2}(1-\psi^2)-
   \frac{(\psi^2)'}{r}=8\pi\rho.
\end{equation*}   
Similarly, combining Eq. (\ref{E:Einstein2}) 
with Eq. (\ref{E:gtt}), yields Eq. (\ref{E:E2}).   


\section{The role of embedding}

Embedding theorems have a long history in
the general theory od relativity.  For example, 
according to Refs. \cite{WP92, SW03}, the 
\emph{vacuum} field equations in five dimensions 
yield the Einstein field equations \emph{with 
matter}, called the \emph{ induced-matter theory},
to be understood in the following sense: what 
we perceive as matter is just the impingement 
of the higher-dimensional space onto ours; this
 may very well include exotic matter.

According to Campbell's theorem \cite{jC26}, 
a Riemannian space can be embedded in a
higher-dimensional flat space: an 
\emph{n}-dimensional Riemannian space is said 
to be of embedding class $m$ if $m+n$ is the
lowest dimension $d$ of the flat space in 
which the given space can be embedded.  Given 
that $d=\frac{1}{2}n(n-1)$, a four-dimensional 
Riemannian space is of class two since it can 
be embedded in a six-dimensional flat space, 
i.e., $d=6$.  Moreover, a line element of class 
two can be reduced to a line element of class 
one by a suitable transformation of coordinates 
\cite{sM19, sM16, MRG17, MGRD17, MM17, MG17}.  
Such a metric can therefore be embedded in the 
five-dimensional flat spacetime
\begin{equation}\label{E:line3}
ds^{2}=-\left(dz^1\right)^2+\left(dz^2\right)^2
+\left(dz^3\right)^2+\left(dz^4\right)^2
+\left(dz^5\right)^2;
\end{equation}
the coordinate transformation is given by
$z^1=\sqrt{K}\,e^{\frac{\nu}{2}}
 \,\text{sinh}{\frac{t}{\sqrt{K}}}$,
  $z^2=\sqrt{K}
 \,e^{\frac{\nu}{2}}\,\text{cosh}{\frac{t}{\sqrt{K}}}$,
 $z^3=r\,\text{sin}\,\theta\,\text{cos}\,\phi$, $z^4=
 r\,\text{sin}\,\theta\,\text{sin}\,\phi$,
 and
 $z^5=r\,\text{cos}\,\theta$.  The differentials of 
 these components are
\begin{equation}\label{E:diff1}
dz^1=\sqrt{K}\,e^{\frac{\nu}{2}}\,\frac{\nu'}{2}\,
\text{sinh}{\frac{t}{\sqrt{K}}}\,dr + e^{\frac{\nu}{2}}\,
\text{cosh}{\frac{t}{\sqrt{K}}}\,dt,
\end{equation}
\begin{equation}\label{E:diff2}
dz^2=\sqrt{K}\,e^{\frac{\nu}{2}}\,\frac{\nu'}{2}\,
\text{cosh}{\frac{t}{\sqrt{K}}}\,dr + e^{\frac{\nu}{2}}\,
\text{sinh}{\frac{t}{\sqrt{K}}}\,dt,
\end{equation}
\begin{equation}
dz^3=\text{sin}\,\theta\,\text{cos}\,\phi\,dr + r\,
\text{cos}\,\theta\,\text{cos}\,\phi\,
d\theta\,-r\,\text{sin}\,\theta\,\text{sin}\,\phi\,d\phi,
\end{equation}

\begin{equation}
dz^4=\text{sin}\,\theta\,\text{sin}\,\phi\,dr + r\,
\text{cos}\,\theta\,\text{sin}\,\phi\,
d\theta\,+r\,\text{sin}\,\theta\,\text{cos}\,\phi\,d\phi,
\end{equation}
and
\begin{equation}
dz^5=\text{cos}\,\theta\,dr\, - r\,\text{sin}\,\theta\,d\theta.
\end{equation} 
The substitution yields
\begin{equation}\label{E:line4}
ds^{2}=-e^{\nu}dt^{2}+\left(\,1+\frac{1}{4}K\,e^{\nu}\,
(\nu')^2\,\right)\,dr^{2}+r^{2}\left(d\theta^{2}
+\sin^{2}\theta\, d\phi^{2} \right).
\end{equation}
Metric (\ref{E:line4}) is therefore equivalent to
metric (\ref{E:line2}) if
\begin{equation}\label{E:lambda}
e^{\lambda}=1+\frac{1}{4}K\,e^{\nu}\,(\nu')^2,
\end{equation}
where $K>0$ is a free parameter.  The result is
a metric of embedding class one.  Eq. (\ref{E:lambda})
can also be obtained from the Karmarkar condition
\cite{kK48}:
\begin{equation}
   R_{1414}=\frac{R_{1212}R_{3434}+R_{1224}R_{1334}}
   {R_{2323}},\quad R_{2323}\neq 0.
\end{equation}
It is interesting to note that Eq. (\ref{E:lambda}) is 
a solution of the differential equation
\begin{equation}
   \frac{\nu'\lambda'}{1-e^{\lambda}}=
   \nu'\lambda'-2\nu''-(\nu')^2,
\end{equation}
which is readily solved by separation of
variables.  So the free parameter $K$ is actually 
an arbitrary constant of integration.   

\section{Thin-shell wormholes}
Our first task in this section is to recall 
from Sec. \ref{S:Introduction} that for a 
Morris-Thorne wormhole, the shape function 
$b=b(r)$ must satisfy the flare-out condition 
$b'(r_0)\le 1$, a geometric requirement that 
can only be satisfied by violating the NEC 
$\rho +p_r<0$.  Our discussion of conformal 
symmetry has yielded Eqs. (\ref{E:gtt}) and 
(\ref{E:grr}).  From Eq. (\ref{E:gtt}), 
$e^{\nu}=Cr^2$, we obtain
\begin{equation}\label{E:nuprime}
   \nu'=\frac{2}{r}.
\end{equation} 
Substituting Eqs. (\ref{E:nuprime}) and 
(\ref{E:grr}) in Eq. (\ref{E:lambda}) from 
the embedding theory, we obtain 
\begin{equation}
   \frac{1}{\psi^2}=1+\frac{1}{4}K
   (Cr^2)\left(\frac{2}{r}\right)^2.
\end{equation}
The result is \begin{equation}\label{E:psi}
   \psi^2=\frac{1}{1+KC}.
\end{equation}
Returning to Eqs. (\ref{E:E1}) and 
(\ref{E:E2}), since $(\psi^2)'=0$,
it follows at once that
\begin{equation}\label{E:nullvectors}
  T_{\alpha\beta}k^{\alpha}k^{\beta}
  =8\pi(\rho+p_r)=0.
\end{equation}
Since the NEC is not violated, we do not get 
a wormhole solution.  We will therefore 
consider instead a thin-shell wormhole by 
first defining a suitable shape function, 
making use of Eq. (\ref{E:psi}):
\begin{equation}\label{E:shape}
   b(r)=r\left(1-\frac{r-r_0}{1+KC}\right).
\end{equation} 
Observe that we have indeed $b(r_0)=r_0$, 
while 
\begin{equation}\label{E:flare}
   0<b'(r_0)=1-\frac{r_0}{1+KC}<1
\end{equation}
for $K$ sufficiently large.  (Recall that 
$C$ was obtained from the junction condition,
Eq. (\ref{E:cutoff})).  Conformally symmetric 
wormholes are also discussed in Ref. 
\cite{pK17}.

A thin-shell wormhole is constructed by taking
two copies of a Schwarzschild spacetime and
removing from each the four-dimensional region
\begin{equation}
   \Omega= \{r\le a\,|\,a>2M\},
\end{equation}
where $a$ is a constant \cite{PV95}.  By
identifying the boundaries, i.e., by letting
\begin{equation}
   \partial\Omega= \{r=a\,|\,a>2M\},
\end{equation}
we obtain a manifold that is geodesically
complete.  In our situation, we take $r=a$ 
to be the cut-off in Eq. (\ref{E:cutoff}) 
since we already know that $b(a)=2M$; 
typically, $a\gg r_0$.

To meet this goal, let us consider the 
surface stresses using the Lanczos equations
\cite{fL04}:
\begin{equation}\label{E:sigma1}
     \sigma=-\frac{1}{4\pi}\kappa^{\theta}_
     {\phantom{\theta}\theta}
\end{equation}
and
\begin{equation}
   \mathcal{P}=\frac{1}{8\pi}(\kappa^{\tau}
   _{\phantom{\tau}\tau}+\kappa^{\theta}
   _{\phantom{\theta}\theta}),
\end{equation}
where $\kappa_{ij}=K^{+}_{ij}-K^{-}_{ij}$ and
$K_{ij}$ is the extrinsic curvature.  Still
following Ref. \cite{fL04},
\begin{equation}
   \kappa^{\theta}_{\phantom{\theta}\theta}=\frac{1}{a}
   \sqrt{1-\frac{2M}{a}}-\frac{1}{a}
      \sqrt{1-\frac{b(a)}{a}}.
\end{equation}
So by Eq. (\ref{E:sigma1}),
\begin{equation}\label{sigma2}
   \sigma=-\frac{1}{4\pi a}
   \left(\sqrt{1-\frac{2M}{a}}-
      \sqrt{1-\frac{b(a)}{a}}\right).
\end{equation}

Given that the shell is infinitely thin, the 
radial pressure is zero.  If the surface density
is denoted by $\sigma$, then the NEC violation 
$\sigma +p_r<0$ implies that $\sigma$ is 
negative, which is completely unphysical.  One 
of the goals in this paper is to show that 
under the assumption of conformal symmetry 
in conjunction with the embedding theory, 
$\sigma$ can be positive.  More precisely, if 
$\mathcal{P}$ denotes the surface pressure, 
then we must have $\sigma +\mathcal{P}<0$ to 
ensure that the NEC is violated on the thin 
shell itself, even though the NEC is met for 
the radial outgoing null vector $(1,1,0,0)$, 
as shown in Inequality (\ref{E:nullvectors}). 
Even though $b(a)=2M$, part of the junction 
formalism is to assume that the junction surface 
$r=a$ is an infinitely thin surface having a 
nonzero density that may be positive or negative.  
For $\sigma$ to be positive, we must have 
$\sqrt{1-\frac{2M}{a}}<\sqrt{1-\frac{b(a)}{a}}$,
which implies that $b(a)<2M$.  So let us assume 
for now that $b(a)\approx 2M$ and return to 
Ref. \cite{fL04}:  

\begin{equation}
   K^{\tau\,+}_{\phantom{\tau}\tau}=\frac{M/a^2}
   {\sqrt{1-2M/a}}
\end{equation}
and
\begin{equation}
   K^{\tau\,-}_{\phantom{\tau}\tau}=
   \frac{1}{2}\nu'(a)\sqrt{1-\frac{b(a)}{a}}.
\end{equation}
Since $\nu'(a)=2/r$ by Eq. (\ref{E:nuprime}), 
the surface pressure is given by
\begin{multline}\label{E:pressure1}
   \mathcal{P}=\frac{1}{8\pi}\left[\frac{M/a^2}
   {\sqrt{1-2M/a}}-\frac{1}{a}
   \sqrt{1-\frac{b(a)}{a}}
   +\frac{1}{a}\sqrt{1-\frac{2M}{a}}
   -\frac{1}{a}\sqrt{1-\frac{b(a)}{a}}\right]\\
   =\frac{1}{8\pi}\frac{1}{\sqrt{1-2M/a}}
   \left(\frac{M}{a^2}-\frac{1}{a}
   \sqrt{1-\frac{b(a)}{a}}
   \sqrt{1-\frac{2M}{a}}\right)\\
   +\frac{1}{8\pi}\left(\frac{1}{a}
   \sqrt{1-\frac{2M}{a}}-\frac{1}{a}
   \sqrt{1-\frac{b(a)}{a}}\right).
\end{multline}
It now becomes apparent that for $b(a)\lesssim 
2M$, the last term on the right-hand side is 
close to zero.  As a result,
\begin{multline}
   \mathcal{P}\approx\frac{1}{8\pi}\frac{1}
   {\sqrt{1-\frac{2M}{a}}}\left[
   \frac{\frac{1}{2}b(a)}{a^2}
   -\frac{1}{a}\left(1-\frac{b(a)}{a}\right)
   \right]\\
   =\frac{1}{8\pi}\frac{1}
   {\sqrt{1-\frac{2M}{a}}}\cdot
   \frac{1}{a}\left(-1+\frac{3}{2}\frac{b(a)}{a}
   \right).
\end{multline}
Using our shape function, Eq. (\ref{E:shape}), 
this leads to
\begin{multline}
   \mathcal{P}\approx\frac{1}{8\pi}
   \frac{1}{\sqrt{1-\frac{2M}{a}}}\cdot\frac{1}{a}
   \left[-1+\frac{3}{2}\left(1-
   \frac{a-r_0}{1+KC}\right)  \right]\\
   =\frac{1}{8\pi}\frac{1}{\sqrt{1-\frac{2M}{a}}}
   \cdot\frac{1}{2a}\left(1-\frac{3(a-r_0)}{1+KC}
   \right).
\end{multline}
We know from the flare-out condition, 
Eq. (\ref{E:flare}), that $1+KC$ is
going to be a fixed quantity.  Moreover, 
$a\gg r_0$; so for $a$ sufficiently large,
$\mathcal{P}$ is negative and bounded away 
from zero, while under the assumption that 
$b\lesssim 2M$, $\sigma$ is close to zero. 
We therefore get $\sigma+\mathcal{P}<0$, 
which was to be shown.

The inequality $\sigma+\mathcal{P}<0$ 
indicates that the NEC has indeed been 
violated on the thin shell.  In a 
Morris-Thorne wormhole, matter that 
violates the NEC is referred to as 
``exotic," a requirement that many 
researchers consider to be unphysical.  
In our situation, however, this violation 
has a physical basis since it is a 
direct consequence of the embedding in 
a higher-dimensional spacetime in 
conjunction with the assumption of 
conformal symmetry.  These properties 
avoid the need to hypothesize the 
existence of the highly problematical 
exotic matter.

\section{Conclusion}
This paper discusses thin-shell wormholes 
based on the standard cut-and-paste technique.  
We assume that the wormhole spacetime admits 
a one-parameter group of conformal motions. 
We also make use of an embedding theorem that 
allows a Riemannian space to be embedded in a 
higher-dimensional flat space.  The extra 
degree of freedom enables us to show that the 
surface energy density $\sigma$ is positive, 
while the surface pressure $\mathcal{P}$ is 
negative, but, in addition, $\sigma+
\mathcal{P}<0$.  So the null energy condition 
has been violated.  For a Morris-Thorne wormhole, 
matter that violates the NEC is referred to as 
``exotic," a condition that many researchers 
consider to be unphysical. In this paper, the 
violation has a physical explanation since it 
is a direct consequence of the enmbedding 
theory in conjunction with the assumption 
of conformal symmetry and can therefore be 
viewed as part of the induced-matter theory.

\end{document}